\documentclass[twocolumn,showpacs,amsmath,amssymb,pre,floatfix]{revtex4}
\usepackage{epsfig,graphicx}
\usepackage{dcolumn}
\usepackage{bm}
\usepackage{ams}

\begin{document}

\title{Doping dependence of spin dynamics of drifting electrons in GaAs bulks}

\author{Stefano Spezia\footnote{Email: stefano.spezia@gmail.com}, Dominique Persano Adorno, Nicola Pizzolato, Bernardo Spagnolo}
 \affiliation{Dipartimento di Fisica e Tecnologie Relative, \\
Universit\`a di Palermo and CNISM-INFM, \\
Viale delle Scienze, edificio 18, I-90128 Palermo, Italy}
\begin{abstract}
We study the effect of the impurity density on lifetimes and
relaxation lengths of electron spins in the presence of a static
electric field in a n-type GaAs bulk. The transport of electrons and
the spin dynamics are simulated by using a semiclassical Monte Carlo
approach, which takes into account the intravalley scattering
mechanisms of warm electrons in the semiconductor material. Spin
relaxation is considered through the D'yakonov-Perel mechanism,
which is the dominant mechanism in III-V semiconductors. The
evolution of spin polarization is analyzed by computing the
lifetimes and depolarization lengths as a function of the doping
density in the range $10^{13}\div5\cdot10^{16}$ $cm^{-3}$, for
different values of the amplitude of the static electric field
 ($0.1 \div 1.0$ kV/cm). We find an increase of the electron
spin lifetime as a function of the doping density, more evident for
lattice temperatures lower than $150$ K. Moreover, at very low
intensities of the driving field, the spin depolarization length
shows a nonmonotonic behaviour with the density. At the room
temperature, the spin lifetimes and depolarization lengths are
nearly independent on the doping density. The underlying physics is
analyzed.
\end{abstract}
%
\pacs{71.70.Ej,72.25.Dc,72.25.Rb}
\maketitle

\section{Introduction}\label{sect1}
Spin dynamics is one of the central focuses of semiconductor
spintronics. In fact, in order to make spintronics an usable
technology, spin signal is required to sustain enough long time
and/or enough long distance during transport. This is necessary to
control and detect the spin polarization in logic operations,
communication and storage of information. The loss of spin
polarization before, during and after the necessary operations is a
crucial point in spin device design. Therefore, a complete
understanding of the effect of temperature, driving field amplitude
and doping density on the spin dephasing is essential
\cite{Flatte09,WuReport2010}.

Until now, the experimental investigation of the doping density
influence on the ultrafast spin dynamics in bulk semiconductors have
been performed at low temperature ($T<80$
K)~\cite{Kikkawa98,Dzhioev2002,Furis2006,Schneider2010,Romer2010}.
In the detailed work of Dzhioev et al.~\cite{Dzhioev2002}, the
dependence of the spin lifetime on the donor concentration at very
low temperatures ($T<5$ K) shows a very unusual behavior,
characterized by the presence of two maxima, ascribed to the
predominance of one of the three different spin-relaxation
mechanisms: hyperfine interaction, anisotropic exchange interaction,
and D'yakonov-Perel (DP) mechanism. Recently, R\"{o}mer et
al.~\cite{Romer2010} have measured the electron-spin relaxation in
bulk GaAs for doping densities close to the metal-to-insulator
transition, finding that at temperatures higher than $30$ K and
densities lower than $8.8\cdot10^{16}$ $cm^{-3}$, all measurements
are consistent with DP spin relaxation of free electrons since all
electrons are delocalized and other spin-relaxation mechanisms can
be neglected.
 From a theoretical point of view, by using a fully microscopic kinetic spin Bloch equation approach,
 Jiang and Wu have predicted
a nonmonotonic dependence of the spin relaxation time on the donor
concentration, showing that the maximum spin relaxation time occurs
at the crossing between the degenerate regime and the non-degenerate
one~\cite{Jiang2009,Shen2009}.

In this work, we study the impurity density effect on the fast
process of relaxation of non equilibrium electron spin polarization
in GaAs bulks, by using a semiclassical Monte Carlo technique to
solve the Boltzmann equation ~\cite{Spezia2010}. We analyze the spin
depolarization of drifting electrons at different lattice
temperatures $T$ by considering only the DP mechanism, which is
dominant in n-type III-V
semiconductors~\cite{WuReport2010,Perel1971}. This mechanism,
effective in the intervals between the collisions, is related to the
spin-orbit splitting of the conduction band in non-centrosymmetric
semiconductors like GaAs~\cite{Perel1971}.

\section{Spin dynamics model and Monte Carlo approach}\label{sect2}
\indent In a semiclassical formalism, the term of the
single-electron Hamiltonian which accounts for the spin-orbit
interaction can be written as
\begin{equation}
H_{SO} = \frac{\hbar}{2}\vec{\sigma}\cdot\vec{\Omega}.
\label{HamiltonianSO}
\end{equation}
It represents the energy of electron spins precessing around an
effective magnetic field [$\vec{B}=\hbar\vec{\Omega}/\mu_Bg$] with
frequency $\vec{\Omega}$, which depends on the orientation of the
electron momentum vector with respect to the crystal
axes~\cite{Perel1971}. The quantum-mechanical description of
electron spin evolution is equivalent to that of a classical
momentum $\vec{S}$ experiencing the effective magnetic field, as
described by the equation of motion
\begin{equation}
\frac{d\vec{S}}{dt}=\vec{\Omega}\times\vec{S}. \label{Poisson}
\end{equation}
Every scattering event reorients the direction of the precession
axis, making the orientation of the effective magnetic field
$\vec{B}$ (that strongly depends on $\vec{k}$) random and
trajectory-dependent, thus leading to spin
dephasing~\cite{Perel1971,Spezia2010}.

The Monte Carlo code used here follows the procedure described in
Ref.~\cite{Persano2000}. The spin polarization vector has been
incorporated into the algorithm as an additional parameter and
calculated for each free carrier, by following the procedure
described in Ref.~\cite{Spezia2010} with the difference that in the
present paper we assume the spin-orbit coupling coefficient in
$\Gamma$-valley ($\beta_{\Gamma}$) equal to $8.2$
$eV\cdot\r{A}^{3}$, as used in Ref.~\cite{Jiang2009} to obtain a
better fit with the
experimental work of Kikkawa and Awschalom~\cite{Kikkawa98}. \\
\indent In our simulations we use a temporal step $\Delta$t of $10$
$fs$ and a $5\cdot 10^4$ electron ensemble to collect spin
statistics. To achieve the steady-state transport regime, we run the
simulation code for a transient time (typically $10^4$ time steps).
After that, all the spin electrons are initialized, the spin
relaxation begins and we collect data. All simulations are performed
in a n-type GaAs bulk with a free electrons concentration varying
into the range $10^{13}\div 5\cdot10^{16}$ $cm^{-3}$ (non-degenerate
regime) by assuming that all donors are ionized. By an exponential
fitting of the decay of the spin polarization we estimate the spin
lifetime $\tau$ and the spin depolarization length
$L$~\cite{Spezia2010}. These parameters satisfy the relation
$L=v_d\cdot\tau$, where $v_d$ is the average drift velocity.
\begin{figure}[htbp]
\centering
\resizebox{0.95\columnwidth}{!}{%
\includegraphics*[height=7cm,width=11cm]{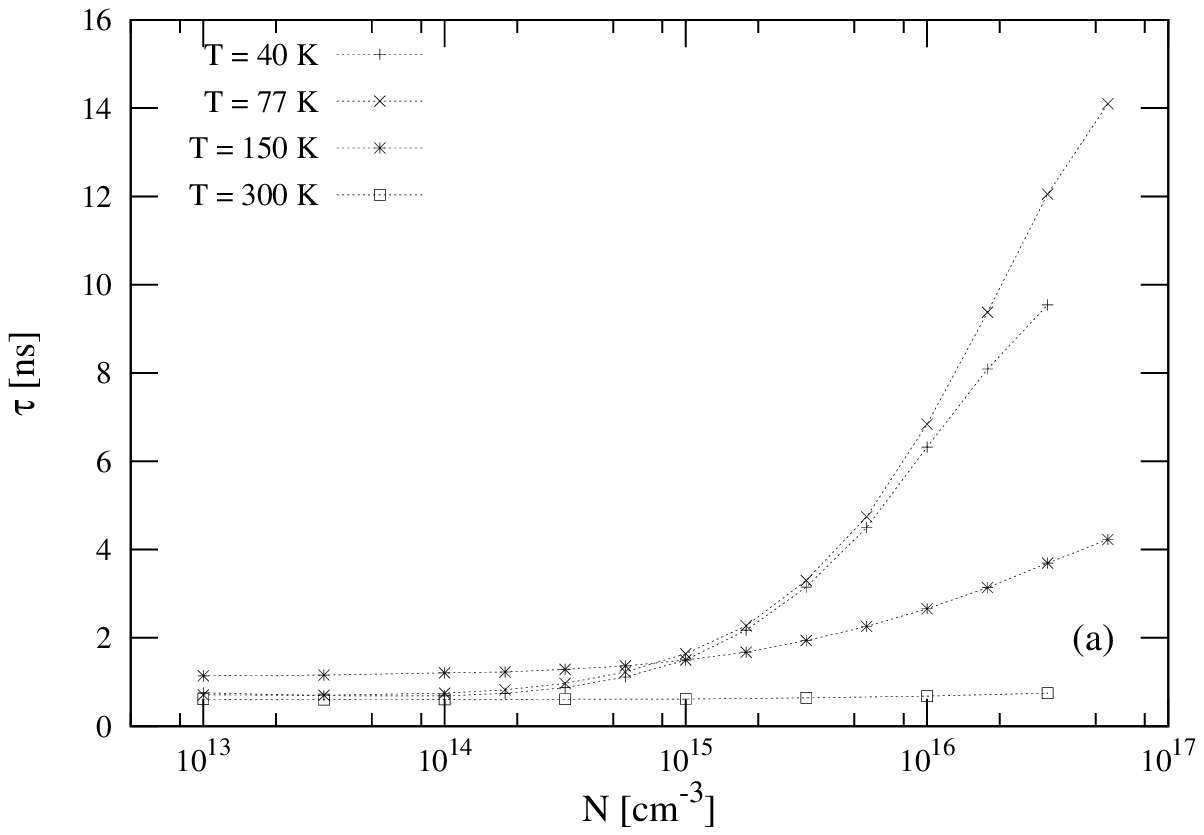}
}
\resizebox{0.95\columnwidth}{!}{%
\includegraphics*[height=7cm,width=11cm]{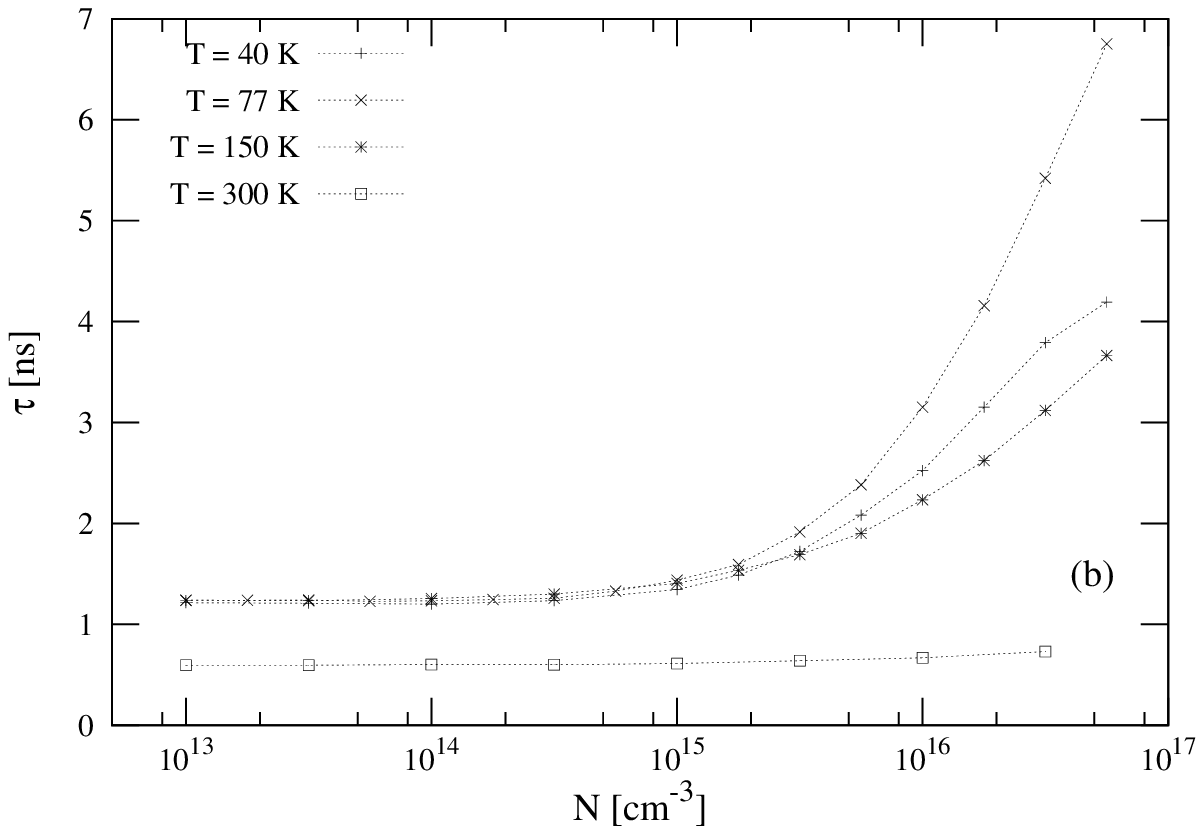}
}
\resizebox{0.95\columnwidth}{!}{%
\includegraphics*[height=7cm,width=11cm]{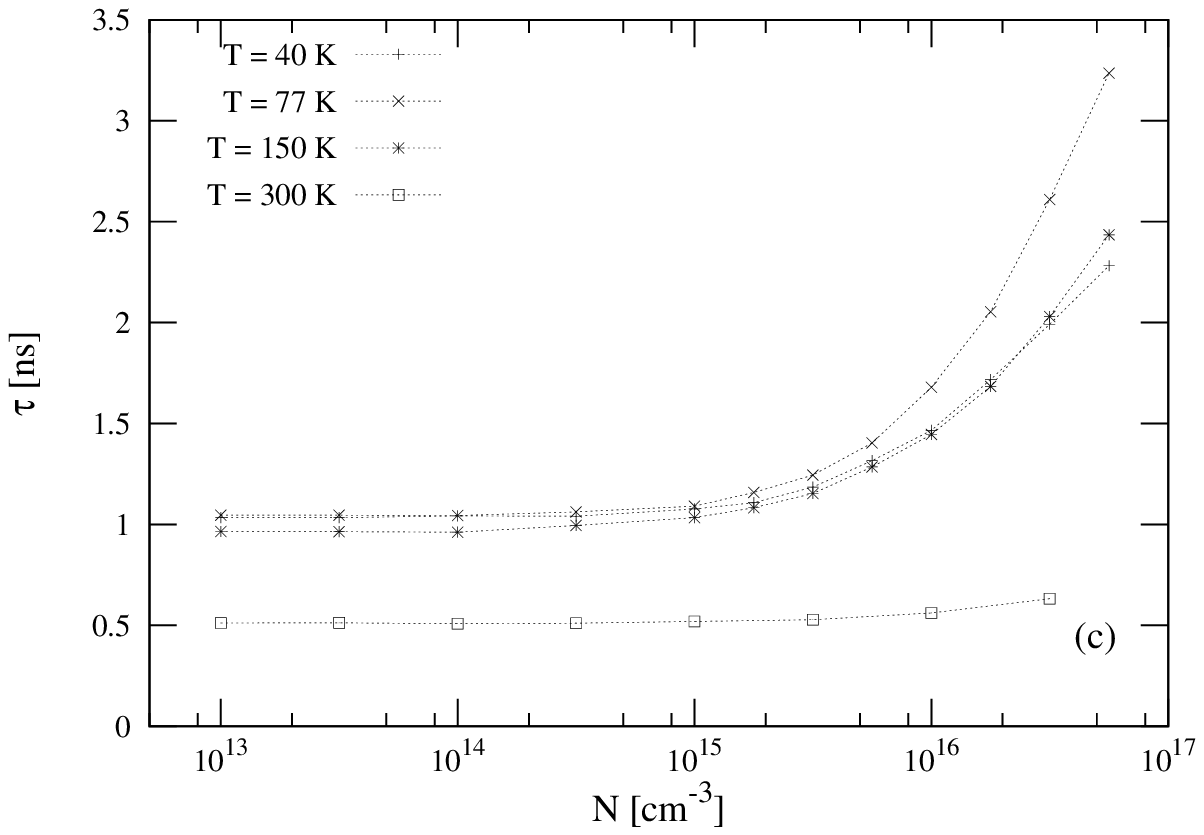}
} \caption{Electron spin lifetimes $\tau$ as a function of the
doping density at different amplitudes of the applied electric field
$F$ $0.1$ kV/cm (a), $0.5$ kV/cm (b) and $1.0$ kV/cm (c), and four
different values of lattice temperature, namely $T=40$, $77$, $150$,
$300$ K.} \label{Fig:1}
\end{figure}
\begin{figure}[htbp]
\centering
\resizebox{0.95\columnwidth}{!}{%
\includegraphics*[height=7cm,width=11cm]{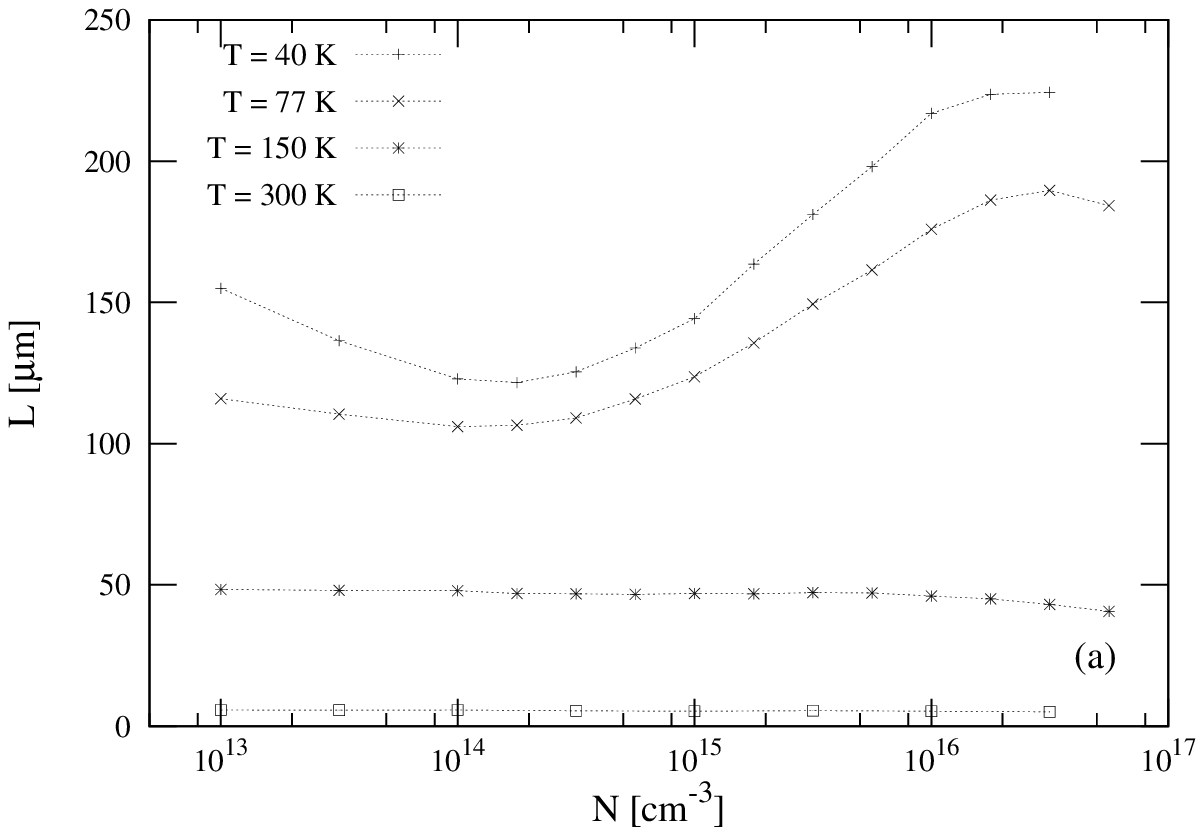}
}
\resizebox{0.95\columnwidth}{!}{%
\includegraphics*[height=7cm,width=11cm]{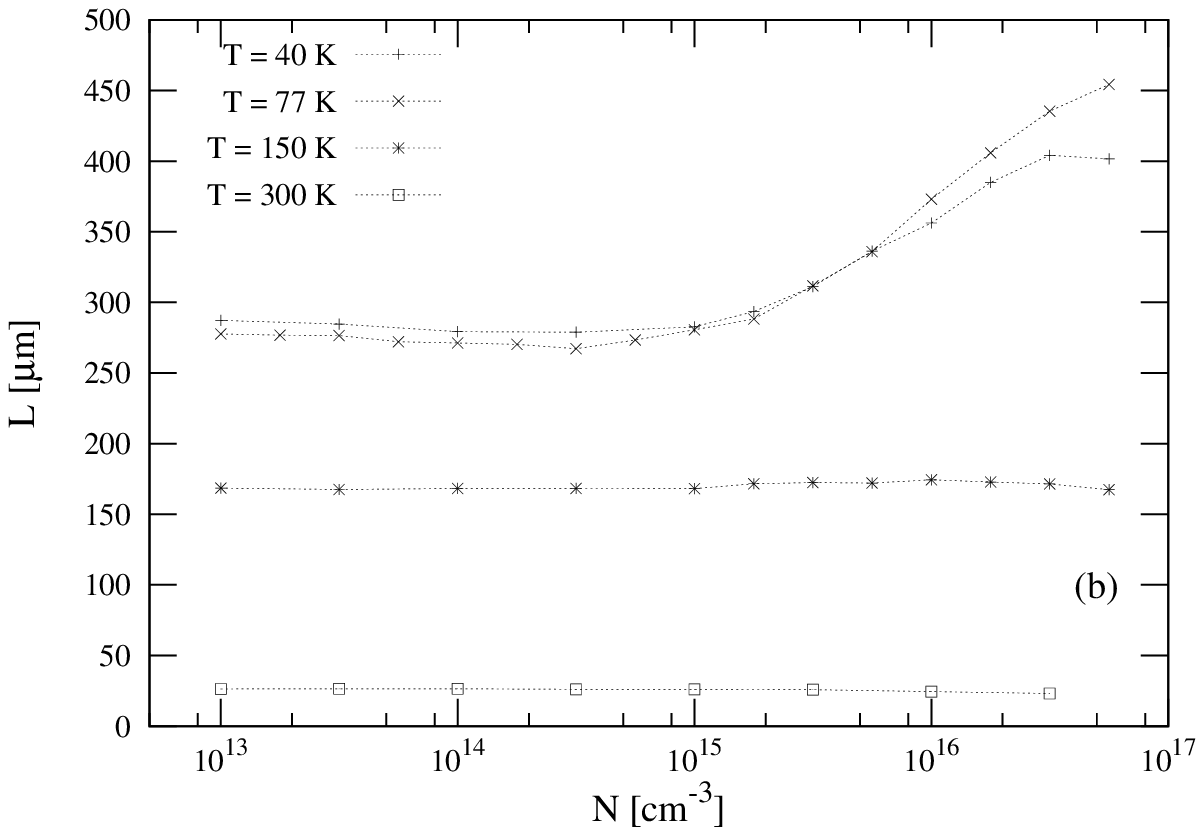}
}
\resizebox{0.95\columnwidth}{!}{%
\includegraphics*[height=7cm,width=11cm]{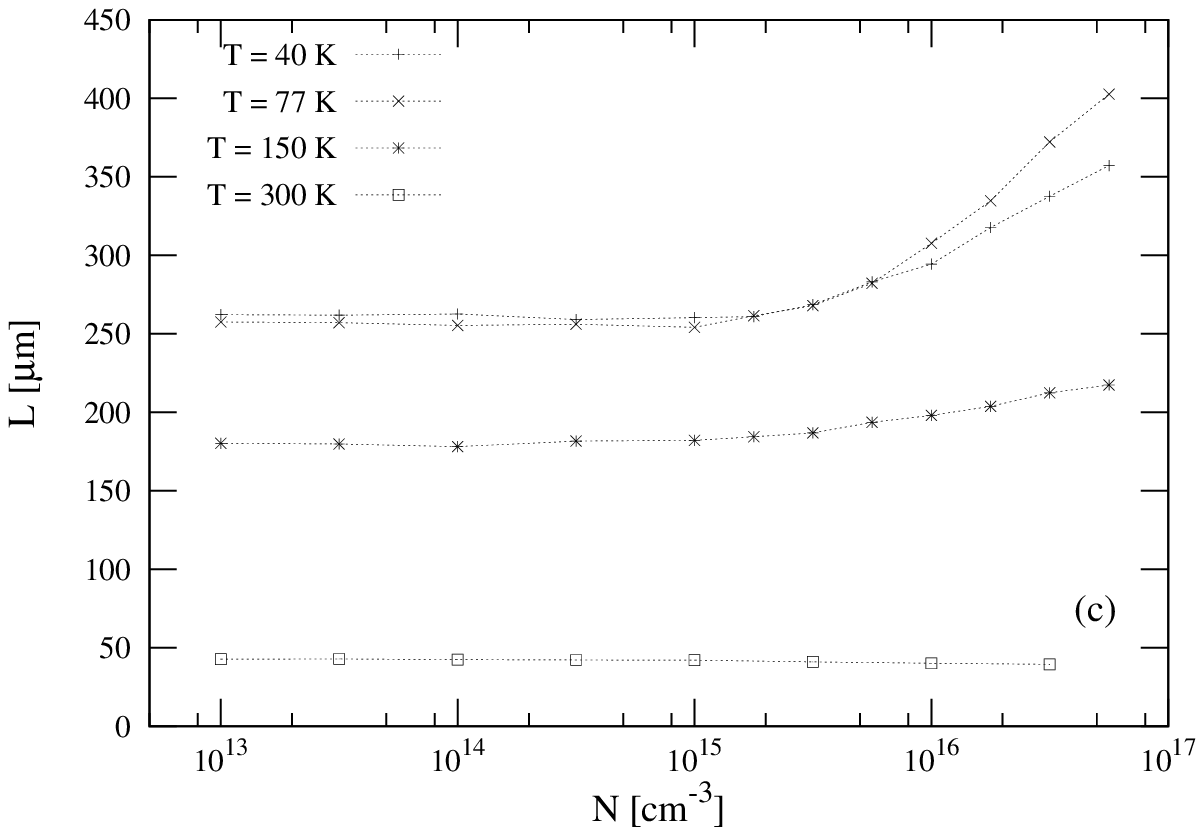}
} \caption{Electron spin depolarization length $L$ as a function of
the doping density at different amplitudes of the applied electric
field $F$ (a) $0.1$ kV/cm, (b) $0.5$ kV/cm and (c) $1.0$ kV/cm, and
four different values of lattice temperature, namely $T=40$, $77$,
$150$, $300$ K.} \label{Fig:2}
\end{figure}

\section{Numerical Results and Discussions}\label{sect3}
In Fig.~\ref{Fig:1}, we show the electron spin lifetime $\tau$ as a
function of the doping density $N$, for different values of applied
electric field $F$, namely $0.1$, $0.5$ and $1.0$ kV/cm. In each
panel, we show four curves at the following values of lattice
temperature $T$: $40$, $77$, $150$ and $300$ K.

Up to $T=150$ K, the electron spin lifetime is nearly independent on
$N$ until $N\sim10^{15}$ $cm^{-3}$, then $\tau$ increases with the
doping density.  For $N>10^{15}$ $cm^{-3}$ and for each value of the
applied field, the longest value of $\tau$ is obtained at $T=77$ K
($\tau_{MAX}\sim14$ ns at $F=0.1$ kV/cm). At the room temperature
($300$ K), the spin lifetimes are almost insensitive to the impurity
density. In the investigated range of $N$, the system is
non-degenerate, i.e. the electron plasma temperature is much greater
than the Fermi's temperature ($T_e\gg T_F$). Hence the inhomogeneous
broadening $<\mid\vec{\Omega}(\vec{k})\mid^2>$ is little sensitive
to $N$, while the momentum scattering rate $\tau_{p}^{-1}$ is
proportional to a linear function of $N$ \cite{Jiang2009}. So, in
accordance with the DP classical relation $\tau\propto
(<\mid\vec{\Omega}(\vec{k})\mid^2>\tau_{p})^{-1}$, for high values
of $N$ the spin lifetime $\tau$ increases with the doping density
\cite{Perel1971}. Moreover, for all the investigated intensities of
the driving field, the relaxation time $\tau$ has a nonmonotonic
behavior as a function of the temperature.

\indent In Fig.~\ref{Fig:2}, we show the electron spin
depolarization length $L$ as a function of the doping density $N$ at
the same values of applied electric fields and lattice temperatures
used in Fig.~\ref{Fig:1}. In particular, in panel (a), i.e. for
$F=0.1$ kV/cm, for $T<150$ K, $L$ appear to be a nonmonotonic
function of $N$, by showing a minimum at $N\sim2\cdot10^{14}$
$cm^{-3}$. At higher temperatures, $L$ is nearly independent on the
doping density. For higher amplitudes of the electric field (panels
(b) and (c)), up to $T= 77$ K, $L$ is insensitive to both the
temperature and the doping density until $N\sim10^{15}$ $cm^{-3}$
and slightly increasing for higher values of $N$. For $T\geq150$ K
the effect of the doping density is marginal. To understand the
behaviour of $L$ as a function of $N$ it is necessary to consider
the interplay between $\tau$ and $v_d$ in the relation
$L=v_d\cdot\tau$. In fact, in the investigated range of $N$, the
spin lifetime $\tau$ always increases with $N$; on the contrary
$v_d$ is a decreasing function of $N$. The nonmonotonic behavior of
$L$, observed at $F=0.1$ kV/cm and $T$ in the range $40\div77$ K,
arises from the fact that for $10^{13}<N<2\cdot10^{14}$ $cm^{-3}$,
$v_d$ decreases more rapidly than $\tau$ increases. Viceversa, for
$N>2\cdot10^{14}$ $cm^{-3}$, $\tau$ increases more quickly and hence
$L$ increases too.

In conclusion, we studied the effect of the doping density on the
ultrafast spin dynamics during drift transport in a GaAs bulk below
the metal-to-insulator transition. For lattice temperatures
$T\leq150$ K, the electron spin lifetime is an increasing function
of the doping density, in accordance with the results reported in
other works using different theoretical
approaches~\cite{Jiang2009,Shen2009} and with the recent
experimental results of R\"{o}mer et al.~\cite{Romer2010}. Moreover,
for very low intensities of the driving field, the spin
depolarization length shows a nonmonotonic behaviour with the
density. At the room temperature, the spin relaxation tends to be
insensitive to the donor concentration.

\section{Acknowledgments} This work was partially supported by MIUR and CNISM-INFM.
The authors acknowledge CASPUR for support by the standard HPC grant
2010.


\end{document}